\documentclass{optica-article} 

\journal{opticajournal} 

\articletype{Research Article} 

\usepackage{braket} 

\newcommand{\ketbra}[2]{\ket{#1}\bra{#2}}
\DeclareMathOperator{\LP}{LP} 

\begin{document}
\title{Spatio-Spectral Quantum State Estimation of Photon Pairs from Optical Fiber Using Stimulated Emission}

\author{Dong Beom Kim\authormark{1,2*} Xiye Hu\authormark{1,2}, Alfred B. U'Ren\authormark{3}, Karina Garay-Palmett\authormark{4}, and Virginia O. Lorenz\authormark{1,2}}

\address{\authormark{1}Department of Physics, University of Illinois Urbana-Champaign, Urbana, Illinois 61801, USA\\
\authormark{2}Illinois Quantum Information Science \& Technology Center (IQUIST), University of Illinois Urbana-Champaign, Urbana, Illinois 61801, USA\\
\authormark{3}Instituto de Ciencias Nucleares, Universidad Nacional Autónoma de México, A.P. 70-543, 04510 Ciudad de México, México\\
\authormark{4}Departamento de Óptica, Centro de Investigación Científica y de Educación Superior de Ensenada, B.C., 22860, Ensenada, México}

\email{\authormark{*}dbkim3@illinois.edu} 


\begin{abstract*}
Developing a quantum light source that carries more than one bit per photon is pivotal for expanding quantum information applications. Characterizing a high-dimensional multiple-degree-of-freedom source at the single-photon level is challenging due to the large parameter space as well as limited emission rates and detection efficiencies. Here, we characterize photon pairs generated in optical fiber in the transverse-mode and frequency degrees of freedom by applying stimulated emission in both degrees of freedom while detecting in one of them at a time. This method may be useful in the quantum state estimation and optimization of various photon-pair source platforms in which complicated correlations across multiple degrees of freedom may be present.
\end{abstract*}


\section{Introduction} \label{sec:Intro}
Developing an efficient quantum light source~\cite{garay2023fiber,anwar2021entangled,krenn2017orbital} that can carry more than one bit of information per photon is crucial for expanding quantum information applications in communication~\cite{willner2021oam,halevi2024high}, computation~\cite{hiekkamaki2021high}, and metrology~\cite{gregory2020imaging,aspden2015photon,moreau2018ghost}. Optical fiber-based photon-pair sources~\cite{garay2023fiber} are an attractive platform that promises easy integration with existing fiber networks and correlations across multiple high-dimensional degrees of freedom (DOF) such as time, frequency, and transverse spatial mode~\cite{cruz2016fiber,garay2016photon,cruz2014configurable}. 

Nevertheless, exploiting such multi-dimensionality requires non-trivial state characterization~\cite{garay2023fiber,zielnicki2018joint,fang2016multidimensional}. This characterization can be challenging to implement with conventional spontaneous-emission measurements including quantum state tomography (QST)~\cite{altepeter2005photonic}. The detection needs to span the entire multi-DOF space~\cite{barreiro2005generation,cruz2016fiber,rambach2021robust}, potentially aided by extended QST methods such as adaptive quantum state tomography~\cite{huszar2012adaptive,mahler2013adaptive}, self-guided tomography~\cite{rambach2021robust,ferrie2014self}, and compressed sensing~\cite{gross2010quantum,bouchard2019compressed}. Moreover, the coincidence-counting measurements involved often require single-photon sensitivity~\cite{bouchard2018measuring,zia2023interferometric}, long integration times~\cite{cruz2014configurable,cruz2016fiber}, and a large number of projective measurements~\cite{barreiro2005generation,rambach2021robust}.

Stimulated-emission tomography (SET)~\cite{liscidini2013stimulated,fang2014fast,rozema2015characterizing,fang2016multidimensional} can speed up characterization through both stimulation and detection in multiple DOFs. The measurements employ classical seed light that stimulates the photon-pair generation process. The higher count rates of the stimulated process lead to more efficient tomography~\cite{liscidini2013stimulated}. These stimulated measurements have previously been applied to a single DOF, such as polarization~\cite{rozema2015characterizing}, frequency~\cite{fang2014fast,liscidini2013stimulated,thekkadath2022measuring}, and transverse spatial mode~\cite{caetano2002conservation,oliveira2021beyond,oliveira2019real,xu2024efficient}, and multiple DOFs including polarization-frequency~\cite{fang2016multidimensional} and polarization-path~\cite{ciampini2019stimulated}.

\begin{figure}[t]
    \centering
    \includegraphics[width=\linewidth]{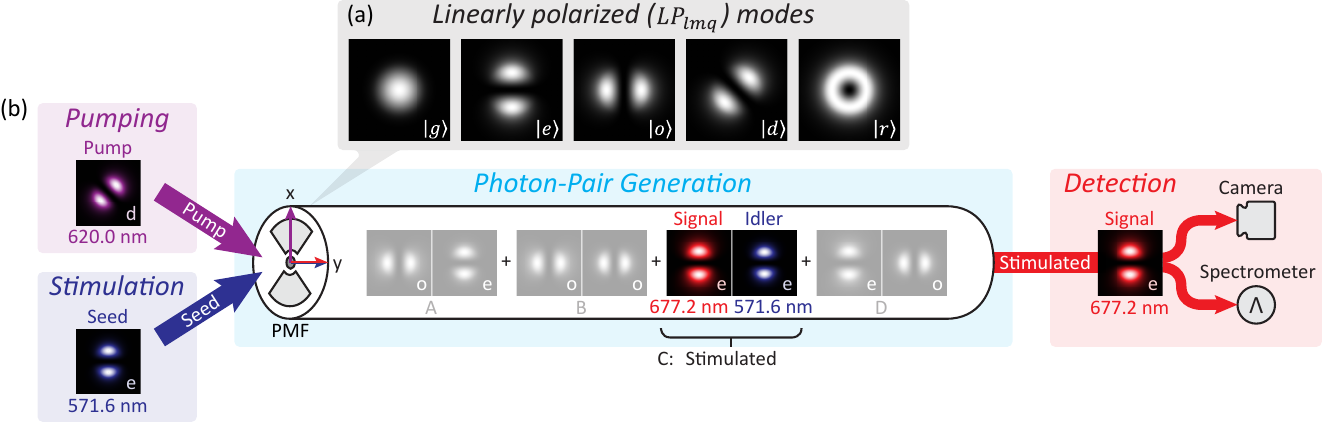}
    \caption{Experimental concept. (a) Intensity distributions of three linearly polarized (LP) modes supported in polarization-maintaining fiber (PMF), $\ket{g=LP_{01}},\ket{e=LP_{11e}}$, and $\ket{o=LP_{11o}}$, and two modes in superposition, $\ket{d}=(\ket{e}+\ket{o})/\sqrt{2}$ and $\ket{r}=(\ket{e}+i\ket{o})/\sqrt{2}$. (b) In our experiment, pump and seed ($=$ idler) in a particular spatio-spectral combination \emph{stimulate} the generation of signal-idler photon pairs in specific modes (indicated with braces) among all observable \emph{spontaneous} four-wave mixing processes labeled A-D (faded out in gray). The stimulated signal photons are measured with a camera or a spectrometer and used to estimate the quantum state of the signal-idler photon pairs.}
    \label{fig:Concept}
\end{figure}

In this work, we extend this effort to introduce a multi-dimensional characterization method that can be applied to sources with correlations in multiple high-dimensional DOFs, in particular transverse spatial mode and frequency. We utilize a few-mode polarization-maintaining fiber source that produces photon pairs correlated in transverse mode and frequency via spontaneous four-wave mixing (SFWM)~\cite{cruz2016fiber,garay2016photon,cruz2014configurable}. We implement stimulated emission in multiple DOFs (transverse mode and frequency), but detect in one DOF at a time (transverse mode or frequency)~\cite{kim2020stimulated}. See Fig.~\ref{fig:Concept} for a graphical representation of the overall experimental concept. We use a seed beam shaped in transverse mode and frequency~\cite{carpenter2012degenerate,carpenter2016complete,fontaine2019laguerre,caetano2002conservation} to stimulate the FWM process, and measure the transverse-mode images and spectra of the stimulated signal using a camera and a spectrometer. Because the transverse modes and spectral modes are correlated, transverse-mode-resolved joint spectral intensities (JSIs) -- an \textit{inter}-DOF -- can be used to investigate the transverse-mode quantum state -- \textit{intra}-DOF -- of the photon pairs. The acquired inter-DOF coherence information can thus yield the intra-DOF coherence information. 

This method reduces the number of measurements while still providing the coherence information across multiple DOFs. Our result also shows that stimulated-emission imaging~\cite{caetano2002conservation,oliveira2019real} can be achieved in fiber platforms, exhibiting a real-time monitoring capability. Consequently, this method can be used in conjunction with quantum state tomography to estimate the quantum state and diagnose the underlying causes of deviations from the target state in each DOF. Our method of extracting multi-dimensional information via stimulated emission using detection in one DOF can immediately aid in optimizing various photon-pair source platforms~\cite{mair2001entanglement,hiekkamaki2021high,feng2019chip,ekici2020graded,sulimany2022all,garay2023fiber,morrison2022frequency} where complicated correlations arise across multiple DOFs and generation processes.

\section{Theory} \label{sec:Theory}
\subsection{Transverse spatial modes in few-mode PMF}
Linearly polarized (LP) modes are the transverse spatial modes supported in a conventional cylindrically-symmetric optical fiber that satisfies the weakly guided approximation~\cite{snyder1983optical}. These modes are denoted as $\LP_{lmq}$, where $l$, $m$, and $q$ are the azimuthal, radial, and parity indices describing their modal structures~\cite{garay2016photon,snyder1983optical}. In this paper, we consider a few-mode polarization-maintaining fiber (PMF) that supports three LP modes: $\ket{\LP_{01}}=\ket{g}$, $\ket{\LP_{11e}}=\ket{e}$, and $\ket{\LP_{11o}}=\ket{o}$ (see Fig.~\ref{fig:Concept}(a)). As transverse-mode basis states, these three LP modes can be combined to form superposition states, e.g., $\ket{d,a}=\left(\ket{e}\pm\ket{o}\right)/\sqrt{2}$ and $\ket{r,l}=\left(\ket{e}\pm i\ket{o}\right)/\sqrt{2}$ as shown in Fig.~\ref{fig:Concept}(a). The modes are then further affected by the two types of birefringence in the PMF: a polarization birefringence $\Delta=n^x-n^y$ between the slow ($x$) and fast ($y$) axes of the PMF and a parity birefringence $\Delta^{p}=n^o-n^e$ between transverse modes with even ($e$) and odd ($o$) parities. Fig.~\ref{fig:Concept} shows how the slow ($x$) axis of the PMF is oriented along the vertical direction and the mode $\ket{e}$ intensity lobes.

When characterizing the photon-pair generation process in a few-mode PMF, it is important to accurately describe the property of a transverse mode at a given wavelength. For this purpose, we define an effective refractive index, which takes into account the transverse geometrical effect of the optical fiber ($T_{\nu}$) as well as its material dispersion property ($\omega_{\nu}$): $n^{T_{\nu}}_{\nu}=n_{\mathrm{eff}}(\omega_{\nu},T_{\nu})$, where $\nu$ indicates pump $p$, signal $s$, or idler $i$ and $T_{\nu}$ and $\omega_{\nu}$ indicate transverse mode and frequency, respectively. Using this convention in the $xx$-$yy$ cross-polarized scheme~\cite{garay2016photon,garay2007photon}, in which the pump is polarized along $x$ and the signal and idler are polarized along $y$, the effective refractive indices of the $\ket{e}$ and $\ket{o}$ modes can be represented as the following: $n_{p}^{ex}=n_{p}^{e}+\Delta$, $n_{p}^{ox}=n_{p}^{ex}+\Delta^p$, 
$n_{s,i}^{ey}=n_{s,i}^{e}$, and $n_{s,i}^{oy}=n_{s,i}^{ey}+\Delta^p$.

\subsection{Four-wave mixing in few-mode PMF}
Utilizing the transverse modes, the few-mode PMF can generate photon pairs correlated in transverse mode and frequency~\cite{cruz2016fiber,garay2016photon,cruz2014configurable} through a nonlinear optical process called spontaneous four-wave mixing (SFWM)~\cite{boyd2020nonlinear}. The SFWM process relies on the $\chi^{(3)}$ nonlinear optical susceptibility of the fiber to annihilate two pump photons ($p_1$, $p_2$) and create a signal ($s$) and an idler ($i$) photon pair.
For this nonlinear process to occur, it needs to satisfy a phase-matching condition, which is determined by the energy ($\Delta\omega=0$) and momentum conservation ($\Delta k=0$) constraints, with
\begin{equation} \label{eq:dk-phasematching}
\begin{split}
    \Delta\omega&=\omega_{p_1}+\omega_{p_2}-\omega_{s}-\omega_{i},\\
    \Delta k&=k_{p_1}+k_{p_2}-k_{s}-k_{i}-k_{NL}\\
    &=n(\omega_{p_1},T_{p_1})\frac{\omega_{p_1}}{c}+n(\omega_{p_2},T_{p_2})\frac{\omega_{p_2}}{c}-n(\omega_{s},T_{s})\frac{\omega_{s}}{c}-n(\omega_{i},T_{i})\frac{\omega_{i}}{c}-k_{NL},
\end{split}    
\end{equation}
where $\omega_{\nu}$, $T_{\nu}$, $k_{\nu}$ are the angular frequency, transverse mode, and wavenumber, respectively of $\nu=\{p_{1},p_{2},s,i\}$, $c$ is the speed of light, and $k_{NL}$ is the nonlinear contribution from self- and cross-phase modulation~\cite{garay2007photon}. Among the different types of SFWM that Eq.~\ref{eq:dk-phasematching} can represent, in this paper, we concentrate on the xx-yy cross-polarized birefringent phase-matching with frequency-degenerate pumps ($\omega_{p}=\omega_{p_1}=\omega_{p2}$) to take advantage of the reduced Raman scattering noise and the number of possible SFWM processes~\cite{garay2007photon,lin2007photon,smith2009photon}. Additionally, since the effective refractive index $n(\omega_{\nu},T_{\nu})$ depends on the transverse mode ($T_{\nu}$) and frequency ($\omega_{\nu}$), the phase-matching condition in Eq.~\ref{eq:dk-phasematching} will vary for different combinations of the two. This can lead to photon pairs in different transverse modes to acquire dissimilar frequencies as shall be shown in Sec.~\ref{sec:Results}.

\subsection{Quantum state representation of photon pairs} \label{sec:Theory-QStRep}
With the fundamentals of the transverse modes and SFWM introduced, we can now express the quantum state $\ket{\psi_{si}}$ of the photon pair created from the few-mode PMF. Assuming cross-polarized birefringent phase-matching and frequency-degenerate pumps, the signal-idler photon pair $\ket{\psi_{si}}$ can be generated in a superposition of $N$ distinct SFWM processes as,
\begin{equation} \label{eq:psi_si}
    \ket{\psi_{si}}=\sum_{j}^{N}\int d\omega_{s}\,d\omega_{i}\,c_{j}\ket{\omega_{s}\omega_{i},T_{s}T_{i},yy,...}_{j}=\sum_{j}^{N}C_{j}\otimes\ket{T_{s}T_{i}}_{j},
\end{equation}
where the prefactors weighting each process $j$ are
\begin{equation} \label{eq:psi_si-coeff}
    c_{j}=M_{cj}\sqrt{P_{p_{1}j}P_{p_{2}j}}f_{j}(\omega_{s},\omega_{i})O_{j}(T_{p_1},T_{p_2},T_{s},T_{i}),\ C_{j}=\int d\omega_{s}\,d\omega_{i}\,c_{j}\ket{\omega_{s}\omega_{i},yy,...}_{j}.
\end{equation}
Here, $\ket{\omega_{s}\omega_{i},T_{s}T_{i},yy,...}_{j}$ represents the signal-idler state from SFWM process $j$ in transverse mode, frequency, polarization, and other implicit degrees of freedom, e.g., position, time, etc. This expression is simplified as $C_{j}\otimes\ket{T_{s}T_{i}}_{j}$ to highlight the transverse-mode contribution. The prefactors $c_{j}$ and $C_{j}$, which determine the relative amplitude and phase of each SFWM process $j$, are functions of average pump power $P_{p_{1,2}j}$, joint spectral amplitude (JSA) $f_{j}(\omega_{s},\omega_{i})$, and transverse-mode overlap integral $O_{j}(T_{p_1},T_{p_2},T_{s},T_{i})$. $O_{j}$ quantifies the spatial overlap of the four transverse modes participating as defined in~\cite{garay2016photon}.  $M_{cj}$ is the normalization constant satisfying $\braket{\psi_{si}|\psi_{si}}=1$.

The JSA $f_{j}(\omega_{s},\omega_{i})$, which contains information about spectral correlations between signal and idler photons for each SFWM process~\cite{garay2016photon,cruz2014configurable,cruz2016fiber,zielnicki2015engineering,zielnicki2018joint,garay2023fiber}, is defined and linearly approximated as~\cite{garay2007photon},
\begin{equation} \label{eq:f-JSA}
    f_{j}(\omega_{s},\omega_{i})=\int d\omega_{p}\,\alpha(\omega_{p})\alpha(\omega_{s}+\omega_{i}-\omega_{p})\phi_{j}(\omega_{s},\omega_{i})\approx\alpha(\omega_{s},\omega_{i})\phi_{j}(\omega_{s},\omega_{i}),
\end{equation}
where $\alpha(\omega_{s},\omega_{i})$ is the pump spectral envelope function and $\phi_{j}(\omega_{s},\omega_{i})$ is the phase-matching function specific for $j$. For degenerate pumps, the JSA can be linearly approximated to $f_{j}(\omega_{s},\omega_{i})\approx\alpha(\omega_{s},\omega_{i})\,\operatorname{sinc}(\frac{L}{2}\Delta k_{j})e^{i\frac{L}{2}\Delta k_{j}}$ where $L$ is the length of the fiber and $\Delta k_{j}$ is the phase mismatch for the process $j$ as defined in Eq.~\ref{eq:dk-phasematching}. For non-degenerate pumps, while the JSA can be linearly approximated in the same form as Eq.~\ref{eq:f-JSA}, it is also a function of the temporal walk-off between the two pumps \cite{fang2013state,zhang2019dual}. In our system, $\alpha(\omega_{s},\omega_{i})$ and $\phi_{j}(\omega_{s},\omega_{i})$ determine the spectral widths of the JSA peak along the diagonal and anti-diagonal directions, respectively. In this paper, we measure the joint spectral intensity (JSI) $|f_{j}(\omega_{s},\omega_{i})|^2$. The joint spectral phase (JSP) is defined as $\arg\{f_{j}(\omega_{s},\omega_{i})\}$. See Supplement~1 for a more comprehensive explanation of the factors in Eq.~\ref{eq:psi_si-coeff} and the quantum state representation of the pump.

The fiber parameters for the PMF considered here (Fibercore HB800C) are obtained through genetic algorithm analysis~\cite{cruz2016fiber} to be: fiber core radius $r=1.74\mathrm{\ \mu m}$, numerical aperture $NA=0.17$, $\Delta=2.37\times10^{-4}$, $\Delta^{p}=4.41\times10^{-4}$. With these parameters and the three transverse modes ($\ket{g}$, $\ket{e}$, and $\ket{o}$) for the pump, signal, and idler, only 10 out of 15 SFWM processes satisfy orbital angular momentum (OAM) and parity conservation and therefore are experimentally realizable~\cite{cruz2016fiber,garay2016photon}. Considering only the $\ket{e}$ and $\ket{o}$ modes, 5 of the above SFWM processes are viable with the following transverse mode combinations $\left(T_{p_1},T_{p_2},T_{s},T_{i}\right)$: A $(e,o,o,e)$, B $(o,o,o,o)$, C $(e,e,e,e)$, D $(e,o,e,o)$, and E $(o,o,e,e)$, where the labels A-E will be used throughout the paper to indicate the corresponding processes.

Using the formalism introduced earlier in Eqs.~\ref{eq:psi_si} and \ref{eq:psi_si-coeff}, these five FWM processes can be obtained with the two pumps in superpositions of $\ket{e}$ and $\ket{o}$ transverse modes, i.e., $\ket{\psi_{p}}=\ket{\psi_{p_1}}=\ket{\psi_{p_2}}=A_{e}\ket{e_{p}}+A_{o}\ket{o_{p}}$ given that we do not have individual control over the two pumps. The quantum states of the pumps $\ket{\psi_{p_{1}p_{2}}}$ and the signal-idler photon pairs $\ket{\psi_{si}}$ can be represented as
\begin{equation} \label{eq:psi_p1p2si-eo}
\begin{split}
    \ket{\psi_{p_{1}p_{2}}}&=\ket{\psi_{p}}^{\otimes2}
    =B_{ee}\ket{e_{p_1}e_{p_2}}+2B_{eo}\ket{e_{p_1}o_{p_2}}+B_{oo}\ket{o_{p_1}o_{p_2}},\\
    \ket{\psi_{si}}&=C_{ee}\ket{e_{s}e_{i}}+C_{eo}\ket{e_{s}o_{i}}+C_{oe}\ket{o_{s}e_{i}}+C_{oo}\ket{o_{s}o_{i}},
\end{split}
\end{equation}
where $A_{j}$ and $B_{j}$ are prefactors similar to $C_{j}$ (see Supplement~1 for details; these are different from the FWM process labels, A, B, and C). Here, $\otimes$ between $A_{j}, B_{j}, C_{j}$ and $\ket{...}_{j}$ are omitted for simplicity. Notice that $B_{eo}\ket{e_{p_1}o_{p_2}}=B_{oe}\ket{o_{p_1}e_{p_2}}$ is satisfied due to their indistinguishability, resulting in an extra factor of 2 before the $B_{eo}$ pump term and the corresponding $C_{eo}$ and $C_{oe}$ signal-idler terms, implicitly through $P_{p_{1,2}j}$ in Eq.~\ref{eq:psi_si-coeff}.

Stimulated emission provides an efficient way to characterize the state in Eq.~\ref{eq:psi_p1p2si-eo} using either the signal or idler as a seed. By controlling the transverse modes and frequencies of the pump and the seed (idler), individual FWM process(es), and thus the photon-pair state in Eq.~\ref{eq:psi_p1p2si-eo} can be selectively excited (see Fig.~\ref{fig:Concept}(b)). This is equivalent to applying a projection operator $\ket{\omega_{i},T_{i},y}_{j_{0}}\bra{\omega_{i},T_{i},y}_{j_{0}}$, which describes the idler of process $j_{0}$, to $\ket{\psi_{si}}$ in Eq.~\ref{eq:psi_p1p2si-eo} to obtain the stimulated photon state $\ket{\omega_{s},T_{s},y}_{j_{0}}$. This process is highly efficient~\cite{fang2016multidimensional,kim2020stimulated} when used in conjunction with a classical seed beam since the stimulated photon number is linearly proportional to the seed photon number~\cite{liscidini2013stimulated,fang2014fast}. This requires sufficiently well-defined pump ($T_{p_1}$, $T_{p_2}$) and seed ($T_{i}$) transverse modes as well as a narrow spectral bandwidth seed ($\omega_{i}$), as will become clear in the following sections.

\section{Methods} \label{sec:Methods}
\begin{figure}[t]
    \centering
    \includegraphics[width=\linewidth]{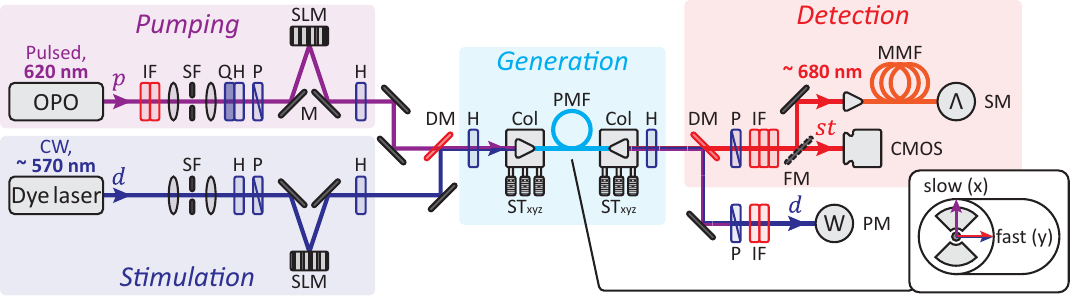}
    \caption{Experimental setup. Spatio-spectrally structured pump and seed beams stimulate a particular photon-pair state in optical fiber that is further detected with a spectrometer or a CMOS camera. The inset shows the slow (x) and fast (y) axes of the PMF. $p$: pump, $d$: seed, $st$: stimulated signal, IF: interference filter, SF: spatial filter consisting of a pinhole and a convex lens pair, Q: quarter-wave plate, H: half-wave plate, P: linear polarizer, M: mirror, SLM: spatial light modulator, DM: dichroic mirror, Col: collimator, $\mathrm{ST_{xyz}}$: xyz-translation stage, PMF: polarization-maintaining fiber, FM: flip mirror, SM ($\Lambda$): spectrometer, PM ($\mathrm{W}$): power meter.}
    \label{fig:ExpSetup}
\end{figure}

We use the experimental setup shown in Fig.~\ref{fig:ExpSetup} to characterize our fiber-based photon-pair source in both transverse mode and wavelength via stimulated emission.
An optical parametric oscillator (OPO, Inspire HF100) generates a pump beam with $\approx$ 200 fs pulse width, 80 MHz repetition rate, 8 mW average beam power, and 620 nm center wavelength. We couple this pump into a 10 cm-long few-mode PMF (HB800C, Fibercore) with its polarization along the slow axis ($x$) for cross-polarized SFWM photon-pair generation as described in Sec.~\ref{sec:Theory}. Continuous wave (CW) ring dye laser (Coherent 899) generates a classical seeded idler beam with narrow 2 GHz linewidth and 2 mW average beam power that is wavelength-tunable around 570 nm. We couple this seed beam into the same PMF with its polarization along the fast axis ($y$) to stimulate the FWM processes.

In order to selectively excite and stimulate specific FWM processes, it is crucial that we precisely control the spatial, spectral, and polarization states of the pump and the seed. We employ reflective phase-only spatial light modulators (SLM, Holoeye Pluto 2), seed laser wavelength calibration with a spectrometer (Andor SR303i with iDus 420), and polarization optics to control the respective DOF. Spatially, we use the SLMs to shape the pump~\cite{boucher2021engineering} to $\ket{d}$ and the seed to $\ket{e}$, $\ket{o}$, and $\ket{d}$ for Sec.~\ref{sec:Results}. In addition to the standard transverse-mode control techniques involving computer-generated SLM phase masks~\cite{carpenter2012degenerate,carpenter2016complete,fontaine2019laguerre,bouchard2018measuring,davis1999encoding,labroille2014efficient}, we adjust the phase mask iteratively until only one FWM process (spectral peak) is observed at a time in the spontaneous or stimulated FWM spectrum. Spectrally, we calibrate the seed laser wavelength scan with a spectrometer (see Supplement~1 for the calibration result). The interference filters spectrally shape the pump to about 2 nm full width at half maximum (FWHM) centered around 620 nm and filter the stimulated signal to a broad (670 to 700) nm range or a narrow $\sim$ 1 nm FWHM range depending on the application. Other optics including wave plates, linear polarizers, and pinhole spatial filters supplement the beam preparation for accurate FWM process excitation. 

For a full characterization of all the FWM processes, we scan the seed wavelength in the (567 to 576) nm range in steps of 0.05 nm. For a given seed wavelength, these stimulated photons are measured in two degrees of freedom, switchable via a flip mirror: wavelength with a spectrometer and transverse mode with a CMOS camera (Thorlabs CS505MU) using $16\mathrm{\ px}\times16\mathrm{\ px}$ pixel binning. A power meter at the PMF output normalizes the measured data with the seed power. These spectral and spatial data are then used to reconstruct the JSI~\cite{fang2014fast} and resolve the corresponding transverse-mode state, respectively, i.e., transverse-mode-resolved JSI.

\section{Results} \label{sec:Results}
\subsection{Transverse-mode-resolved JSI} \label{sec:TModeResJSI}
\begin{figure}[ht!]
    \centering
    \includegraphics[width=0.8\linewidth]{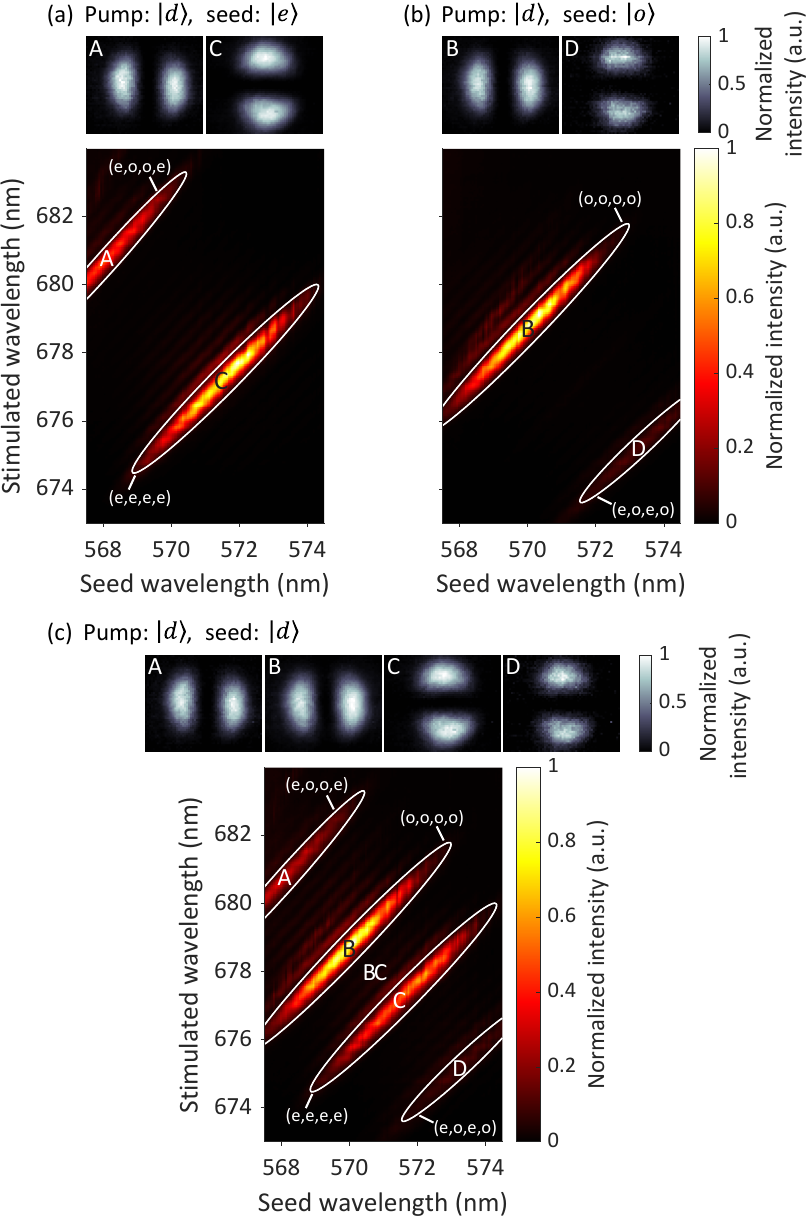}
    \caption{Transverse-mode images (top) of the stimulated signal and JSI plots (bottom) of the FWM processes for different seed transverse modes, (a) $\ket{e}$, (b) $\ket{o}$, and (c) $\ket{d}$. The pump transverse mode is fixed to $\ket{d}$. Each FWM process (A-D) with transverse modes $(T_{p_1},T_{p_2},T_{s},T_{i})$ is specified with a solid $1/e^{3}$ two-dimensional Gaussian fit contour. The transverse mode images are captured with camera exposure times of (a-b) 200 ms and (c) 400 ms at the JSI peaks with narrow spectral filters on the stimulated signal to block other FWM contributions. The intensities of all the transverse mode images and JSI plots are normalized to one. See Visualization~1 and 2 for a real-time video and an animated version of this data.}
    \label{fig:JSI,Images}
\end{figure}
Using the methods described in Sec.~\ref{sec:Methods}, we measure transverse-mode-resolved JSIs. Figure~\ref{fig:JSI,Images} shows the measured JSI plots of the photon pairs and the transverse-mode images of the stimulated signal photons. The signal transverse modes are imaged with exposure times of 200 ms for Fig.~\ref{fig:JSI,Images}(a-b) and 400 ms for Fig.~\ref{fig:JSI,Images}(c). Narrow signal spectral filters are used to isolate individual FWM processes. Note that the states of the stimulated signal ($st$) and the seed ($d$) will reflect those of the spontaneously generated signal ($s$, around 680 nm) and the idler ($i$, around 570 nm) photons, respectively.

We vary the seed transverse modes to $\ket{e}$, $\ket{o}$, and $\ket{d}$ (see Fig.~\ref{fig:JSI,Images}(a-c)) while keeping the pump mode at $\ket{d}=(\ket{e}+\ket{o})/\sqrt{2}$. This choice of seed transverse mode as well as its wavelength changes the JSI and the stimulated signal transverse mode, which helps isolate different FWM processes. Specifically, only the FWM processes that involve the given idler (seed) transverse mode and wavelength manifest in the JSI plot and the signal images. For example, with $\ket{e}$ ($\ket{o}$) seed and $\ket{d}$ pump, only the A and C (B and D) processes are stimulated, as shown in Fig.~\ref{fig:JSI,Images}(a) (Fig.~\ref{fig:JSI,Images}(b)). On the other hand, with the seed in superposition state $\ket{d}$, all of the four FWM processes are stimulated, as shown in Fig.~\ref{fig:JSI,Images}(c) (process E is outside our spectral range of interest and expected to appear at $(\lambda_{s},\lambda_{i})_{E}\sim(730,540)_{E}\mathrm{\ nm}$). Through further measurements with $\ket{e}$ and $\ket{o}$ pumps to resolve the remaining ambiguity in the pump transverse modes, we can conclude that each JSI lobe is associated with a $\left(T_{p_1},T_{p_2},T_{s},T_{i}\right)$ FWM process as labeled in Fig.~\ref{fig:JSI,Images}: A $(e,o,o,e)$, B $(o,o,o,o)$, C $(e,e,e,e)$, and D $(e,o,e,o)$. To determine the center signal and idler wavelengths $(\lambda_s,\lambda_i)_{j}$ of the corresponding process, we fit each JSI lobe with a Gaussian function giving: $(680.7, 568.1)_A\textrm{\ nm}$, $(678.7, 570.0)_B\textrm{\ nm}$, $(677.2, 571.6)_C\textrm{\ nm}$, and $(675.3, 573.3)_D\textrm{\ nm}$. Remarkably, this characterization is also possible in real time (see Visualization~1 and 2), similar to~\cite{oliveira2019real} with free-space nonlinear crystals. Supplement~1 provides more information on the characterization efficiency and the relative intensities of the JSI lobes and their relation to the transverse-mode overlap integral $O_j$.

This characterization capability is instrumental in assessing the degree of spectral overlap among different FWM processes, which is essential for creating transverse-mode entanglement as we investigate in Sec.~\ref{sec:QStEstimation}. The FWM processes are spectrally separated in the JSI due to different phase matching conditions and effective refractive indices of transverse modes, as explained in Sec.~\ref{sec:Theory}. Consequently, this means that the quantum state of a signal-idler photon pair expressed in the transverse-spectral-mode basis, $\ket{\psi_{si}}=\sum_{j}\int d\lambda_s\,d\lambda_i\,c_j\ket{T_{s}T_{i},\lambda_s\lambda_i}_{j}$ where $j$ denotes a FWM process, becomes a mixed state in the transverse-mode basis with the spectral DOF is traced out, i.e., a reduced density matrix, $\rho_{si}^T=\operatorname{tr}_{\lambda}(\rho_{si})=\sum_{j}C_{j}\ket{T_{s}T_{i}}_{j}\bra{T_{s}T_{i}}_{j}$.

\begin{figure}[ht!]
    \centering
    \includegraphics[width=0.9\linewidth]{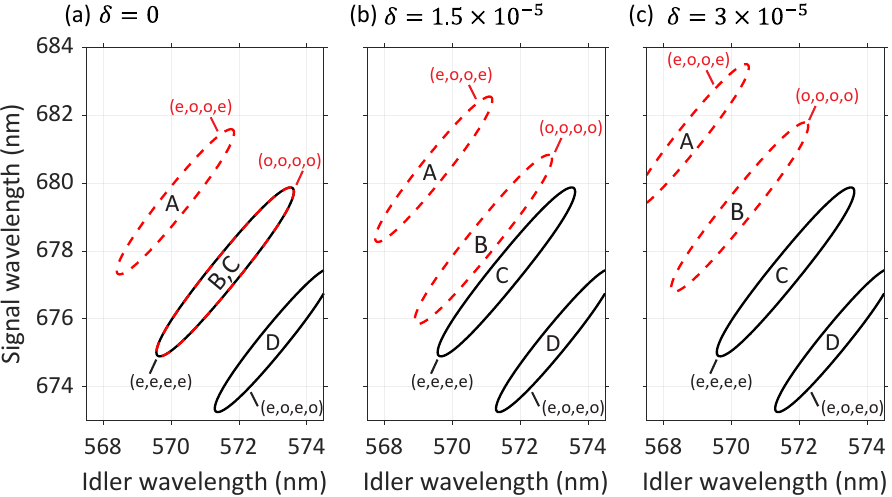}
    \caption{Numerically simulated JSI plots for varying parity birefringence dispersion, $\delta$: (a) $0$, (b) $1.5\times10^{-5}$, and (c) $3\times10^{-5}$. The solid and dashed $1/e^3$ contours correspond to the SFWM processes labeled (A-D) and $(T_{p_1},T_{p_2},T_{s},T_{i})$. Increasing the parity birefringence dispersion $\delta$ increases the spectral separation between C (e,e,e,e) and B (o,o,o,o) by translating A (e,o,o,e) and B (o,o,o,o) (dashed) towards the top left. The pump is fixed to $\ket{d}$.}
    \label{fig:JSINumSim}
\end{figure}

We now turn our attention to the processes B and C. We discovered that in order to explain the spectral separation between B and C, apparent in Fig.~\ref{fig:JSI,Images}(c) and consistently observed in previous studies~\cite{kim2023generating,kim2023towards}, a new correction parameter called parity birefringence dispersion $\delta$ needs to be introduced. Without such correction, the numerical simulation may incorrectly predict the two FWM processes to completely overlap in JSI (see Fig.~\ref{fig:JSINumSim}(a)), which is instrumental for enabling transverse-mode entanglement~\cite{ekici2020graded,kim2023generating,kim2023towards}. For simplicity, here we assume that $\delta$ is a constant describing the difference between the signal and the idler parity birefringences, i.e., $\delta=\Delta^p_s-\Delta^p_i$. With numerical simulation, we vary $\delta$ and observe the change in JSI, as shown in Fig.~\ref{fig:JSINumSim}. We find the non-zero dispersion of $\delta\sim3\times10^{-5}$ (see Fig.~\ref{fig:JSINumSim}(c)) best explains the experimental results presented in Fig.~\ref{fig:JSI,Images}(c). Although $\delta$ is an order of magnitude smaller than $\Delta^p$ ($\sim10^{-4}$), it contributes significantly to the spectral distinguishability between B and C processes, as shown in Fig.~\ref{fig:JSINumSim}. The remaining discrepancies between the experimental data and numerical simulation arise due to imperfect estimation of the fiber parameters. Therefore, more accurate analysis using a full genetic algorithm calculation~\cite{garay2016photon,cruz2016fiber} along with precise measurement of $\delta$ as a function of wavelength may help improve the agreement.

\subsection{Transverse-mode quantum state estimation from transverse-mode-resolved JSIs} \label{sec:QStEstimation}
Building upon the characterization results described in Sec.~\ref{sec:TModeResJSI}, we apply stimulated emission to characterize a fiber source that generates photon pairs with partial transverse-mode entanglement. This characterization allows us to numerically estimate the quantum state of the photon pairs created in the transverse-mode basis. Through this process, we identify potential factors that can degrade the transverse-mode entanglement and find ways to optimize the source accordingly.

To create a maximally entangled transverse-mode Bell state in our system~\cite{kim2023generating,kim2023towards}, $\ket{\psi_{si}}=\left(\ket{e_{s}e_{i}}+\ket{o_{s}o_{i}}\right)/\sqrt{2}$, indistinguishabilities between $\ket{e_{s}e_{i}}$ and $\ket{o_{s}o_{i}}$ in all other degrees of freedom are necessary, i.e., $C_{ee}=mC_{oo}$ (see Eq.~\ref{eq:psi_si-coeff}), where $m$ is a constant. Consequently, the JSIs (Eq.~\ref{eq:f-JSA}) need to satisfy the following overlap condition at the desired frequencies, $\omega_{s}$ and $\omega_{i}$: $\left|f_{ee}(\omega_{s},\omega_{i})\right|^2=\left|f_{oo}(\omega_{s},\omega_{i})\right|^2$. For this, we employ a shorter cross-spliced PMF ($2.5\mathrm{\ cm}\times2$, HB800C) with the same experimental setup used in Sec.~\ref{sec:Methods}. The smaller the fiber length $L$, the wider the spectral bandwidth of the phase-matching function $\phi(\omega_{s},\omega_{i})$ (see Sec.~\ref{sec:Theory-QStRep}) and thus the more spectral overlap arise along the anti-diagonal direction in the JSI. Cross-splicing, where we fusion splice two 2.5 cm-long PMFs such that the second fiber's slow axis is aligned along the first fiber's fast axis, helps compensate for temporal walk-off between the $\ket{e}$ and $\ket{o}$ modes~\cite{meyer2013generating}.

\begin{figure}[t!]
    \centering
    \includegraphics[width=\linewidth]{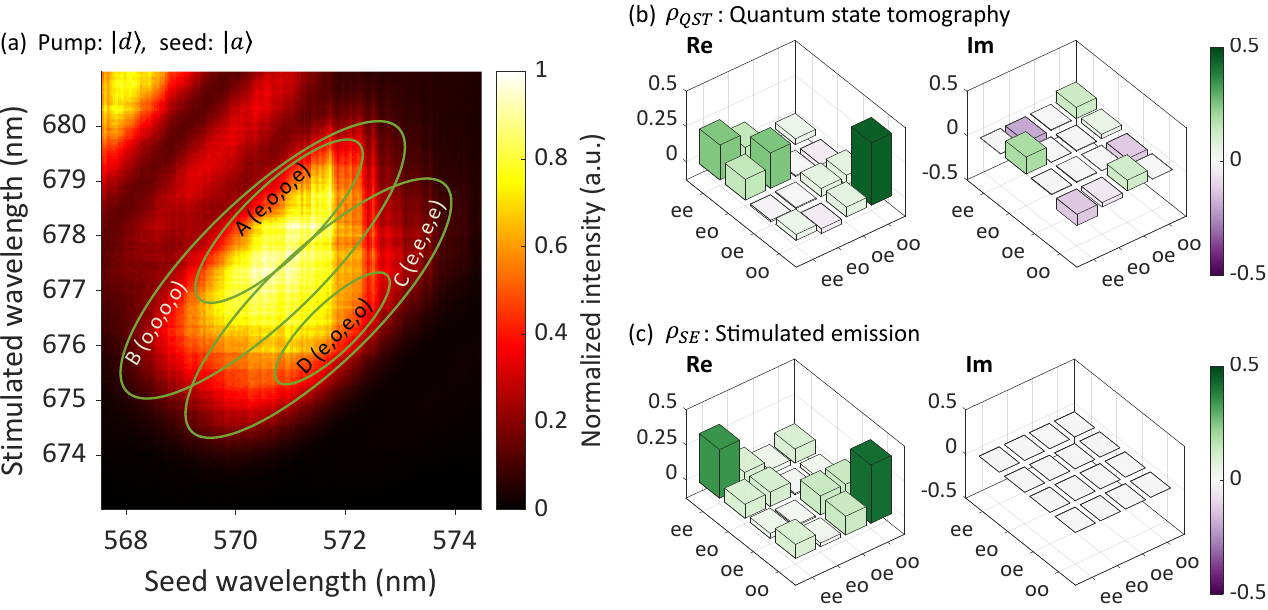}
    \caption{(a) JSI plot of a ($2.5\mathrm{\ cm}\times2$) cross-spliced PMF with the pump in $\ket{d}$ and the seed in $\ket{a}$. $1/e^2$ contours (solid lines) show the two-dimensional Gaussian fits for FWM processes $(T_{p_1},T_{p_2},T_{s},T_{i})$. (b-c) Density matrices describing the transverse-mode quantum state of the photon pairs created: (b) $\rho_{QST}$ measured with a transverse-mode quantum state tomography (QST) and (c) $\rho_{SE}$ estimated from the stimulated-emission measurements (a). The fidelity $F$ between the two density matrices is 0.73, which increases to 0.85 when disregarding the phase.}
    \label{fig:QSTVsSE}
\end{figure}

With a shorter fiber for spectral indistinguishability and cross-splicing for temporal indistinguishability, we measure transverse-mode-resolved JSIs as in Sec.~\ref{sec:Methods}. To identify the FWM processes, similar to Fig.~\ref{fig:JSI,Images}, we conduct a series of measurements with 5 different pump-seed transverse-mode combinations ($e$-$e$, $o$-$o$, $d$-$e$, $d$-$o$, and $d$-$a$). Figure~\ref{fig:QSTVsSE}(a) shows the measured JSI with the pump in $\ket{d}$ and the seed in $\ket{a}$, where we have labeled the spectral peaks with the associated transverse modes $(T_{p_1},T_{p_2},T_{s},T_{i})$ and FWM processes as before. The solid curves in Fig.~\ref{fig:QSTVsSE}(a) again represent the $1/e^{2}$ contours of the two-dimensional Gaussian curve fittings (goodness of fit $R^2\approx0.9$). The B and C contours exhibit some spectral overlap, promising some transverse-mode entanglement at the intersection. Compared to Fig.~\ref{fig:JSI,Images}(a-c), in Fig.~\ref{fig:QSTVsSE}(a), the four processes A, B, C, and D are positioned much closer despite considering the fiber length change effect -- A and D are located fully inside B and C. Here, we attribute this deviation to the difference in fiber parameters. Even for the same type of fiber, HB800C, the fiber parameters can vary spool to spool, which may result in different FWM peak positions. Here, given that the fibers used for Figs.~\ref{fig:JSI,Images}(a-c) and ~\ref{fig:QSTVsSE}(a) are from different spools, we identify that this shorter PMF likely has a smaller parity birefringence $\Delta^{p}\sim1\times10^{-4}$ compared to that used in Sec.~\ref{sec:TModeResJSI}.

Before estimating the signal-idler quantum state with our stimulated emission approach, for reference, we measure the state using a conventional transverse-mode quantum state tomography (QST)~\cite{altepeter2005photonic}. Specifically, we project the signal-idler transverse-mode states into six mutually unbiased measurement basis states ($e, o, d, a, r, l$) and conduct 36 coincidence measurements ($ee, eo, ..., lr, ll$)~\cite{altepeter2005photonic,mair2001entanglement,agnew2011tomography,langford2004measuring,hiekkamaki2019near,kim2023towards} by installing an additional SLM, single-mode fibers, single-photon detectors, and a coincidence counter in the \emph{detection} part of the setup in Fig.~\ref{fig:ExpSetup} (see Supplement~1 for more details). Figure~\ref{fig:QSTVsSE}(b) shows the measured density matrix $\rho_{QST}$ with partial transverse-mode entanglement quantified by concurrence $=0.27\pm0.03$. It has a fidelity, or closeness, to the target Bell state of $0.48\pm0.02$ and a purity of $0.52\pm0.01$. As with typical quantum state tomography results, we can only roughly ascribe the low concurrence, fidelity, and purity to low $\ketbra{ee}{oo}$ and $\ketbra{oo}{ee}$ off-diagonal cross terms and non-zero $\ketbra{oe}{oe}$ and $\ketbra{eo}{eo}$ on-diagonal components each describing the coherence and purity of $\ket{ee}$ and $\ket{oo}$ states, respectively. We can hypothesize the origins of such contributions, but it will be challenging to trace and verify them experimentally without conducting additional measurements. The errors presented here are computed from $10^2$ randomly sampled density matrices assuming Poissonian noise in the coincidence counts for QST.

To compare with the QST, we estimate the transverse-mode density matrix $\rho_{SE}$ from the transverse-mode-resolved JSI, which can provide additional information about the sources of low concurrence, fidelity and purity. We start the estimation procedure by characterizing the transverse-mode density matrix $\rho_{tot}(\lambda_{s},\lambda_{i})$ ($\rho_{tot}(\omega_{s},\omega_{i})$) in the signal-idler wavelength (frequency) space. Using spectral decomposition~\cite{nielsen_chuang_2010}, the full density matrix at given signal and idler wavelengths can be represented as a linear combination of density matrices $\rho_{j}$ as $\rho_{tot}(\lambda_{s},\lambda_{i})=\sum_{j}m_{j}\rho_{j}(\lambda_{s},\lambda_{i})$, where each $\rho_{j}=\ketbra{\psi_{j}}{\psi_{j}}$ describes a pure quantum state $\ket{\psi_j}$ of a FWM process $j$ satisfying the normalization condition $\sum_{j}m_{j}=1$ with $m_{j}\geq0$. For example, $\rho_{B}$, $\rho_{C}$, and $\rho_{B\cap C}$ each represents a pure signal-idler state within the boundary of the process B, C, and the intersection of B and C, respectively, where $\ket{\psi_{B}}=C_{B}\otimes\ket{oo}_B$, $\ket{\psi_{C}}=C_{C}\otimes\ket{ee}_C$, and $\ket{\psi_{B\cap C}}=C_{C}\otimes\ket{ee}_{C}+C_{B}\otimes\ket{oo}_{B}$. In general, if  $N$-FWM processes exist, there will be $2^{N}-1$ binary combinations of $\rho_{j}$'s that specify whether a given signal-idler wavelength coordinate lies inside or outside of a certain FWM process. Representing each FWM process with a two-dimensional Gaussian fitting function as shown in Fig.~\ref{fig:QSTVsSE}(a) simplifies the density matrix calculation at a given signal-idler wavelength domain and thus the estimation procedure. Then, we integrate all these point-wise density matrices in the given signal-idler spectral range experimentally defined by interference filters, thereby producing a single transverse-mode density matrix we name $\rho_{SE}$. In other words, $\rho_{SE}$ is a reduced density matrix $\rho_{tot}^{T}$ in the transverse-mode domain, where the spectral DOF is traced out: $\rho_{SE}=\rho_{tot}^{T}=tr_{\lambda}(\rho_{tot}(\lambda_{s},\lambda_{i}))=\int d\lambda_{s}\,d\lambda_{i}\,\rho_{tot}(\lambda_{s},\lambda_{i})$.

Figure~\ref{fig:QSTVsSE}(c) shows the stimulated-emission estimated density matrix $\rho_{SE}$ using this calculation accounting for the interference filters used in the QST measurement (the full range shown in Fig.~\ref{fig:QSTVsSE}(a); $\lambda_{i}=[567.5, 574.5]\mathrm{\ nm}, \lambda_{s}=[673.0, 681.0]\mathrm{\ nm}$). $\rho_{SE}$ exhibits concurrence $=0.00$, Bell fidelity $=0.48$, and purity $=0.40$. Except for the relative amplitude of $\ketbra{ee}{ee}$ and $\ketbra{eo}{eo}$ elements, overall, $\rho_{SE}$ shows a similar trend as the quantum state tomography result, $\rho_{QST}$  -- high $\ketbra{ee}{ee}$ and $\ketbra{oo}{oo}$, non-zero $\ketbra{ee}{oo}$ and $\ketbra{oo}{ee}$ coherent interaction elements, and other residual elements. Quantitatively, the fidelity $F$ that describes the degree of similarity between the QST and stimulated-emission estimated states is $F(\rho_{QST},\rho_{SE})=0.73$. $\rho_{SE}$ provides a closer estimation if the phase information can be ignored, i.e., $F(|\rho_{QST}|,\rho_{SE})=0.85$. This is related to the current phase measurement limitation of our method as shall be discussed later in Sec.~\ref{sec:Discussion}. 

Despite the remaining discrepancies, $\rho_{SE}$ can still provide sufficient information to deduce potential factors leading to low transverse-mode entanglement, namely, the presence of spectral distinguishabilities among the FWM processes. Within the given spectral window in Fig.~\ref{fig:QSTVsSE}(a), imperfect spectral overlap between the processes B and C can be observed, as well as the presence of other processes A and D. Based on the previous discussions, we can realize that in the transverse-mode basis, spectral overlap between the two processes gives coherence (off-diagonal components in the density matrix), whereas spectral separation gives incoherence (no off-diagonals, leading to a mixed state). Therefore, between $\ket{oo}$ (B) and $\ket{ee}$ (C), we can logically predict that the density matrix will have slight off-diagonal coherence from the B-C overlap in the JSI, as well as the mostly on-diagonal incoherence from the remaining spectrally non-overlapping regions. Similarly, examining the JSI plot in Fig.~\ref{fig:QSTVsSE}(a), we can infer that while $\ket{eo}$ (D) and $\ket{oe}$ (A) will not have off-diagonal elements, they will have non-zero off-diagonal values with $\ket{oo}$ (B) and $\ket{ee}$ (C), respectively (see Fig.~\ref{fig:QSTVsSE}(c)). Ultimately, all these factors contribute to low transverse-mode entanglement. As such, the stimulated-emission method can help probe spectral distinguishabilities that are challenging to assess solely based on the transverse-mode QST result.

Considering the spectral origin, we may now think about tailored strategies to optimize the source for higher transverse-mode entanglement, that is, lowering the spectral indistinguishability. In principle, we can do so by choosing a narrower spectral window that focuses on the B-C intersection area at the cost of reduced counts. Since we can choose an arbitrary spectral window when calculating $\rho_{SE}$, we can easily simulate to determine the optimal filtering strategy. For example, making a $1.5\mathrm{\ cm}\times2$ fiber source out of the Sec.~\ref{sec:TModeResJSI} fiber and using a square-shaped 1 nm-wide spectral window centered at B-C intersection can produce a photon-pair state with concurrence = 0.82, Bell fidelity = 0.91, and purity = 0.84. More fundamentally, we may engineer the phase-matching ~\cite{liu2022engineering} to increase the B-C overlap while decreasing the A and D contributions.

\subsection{Discussion on possible improvements} \label{sec:Discussion}
The aforementioned difference in quantitative measures (concurrence, Bell fidelity, and purity) between $\rho_{QST}$ and $\rho_{SE}$ can be explained by the following assumptions made in the calculation, which may improve with adequate treatments. First, we assume a Gaussian phase-matching function instead of the more accurate sinc (degenerate pumps) and complex error functions (non-degenerate pumps) described in Sec.~\ref{sec:Theory-QStRep}. A Gaussian function can underestimate the spectral overlap that may originate from the tails of the sinc and non-degenerate pump functions~\cite{garay2007photon}, reducing the overall entanglement. As a resolution, we may fit each FWM process with an accurate phase-matching function model accounting for the pump degeneracy (B, C: degenerate, A, D: non-degenerate). Alternatively, we may directly use a non-fitted raw JSI data that could be measured by exciting only one FWM process at a time using a precise transverse-mode control. Second, we assume a flat joint spectral phase (JSP) for all the FWM processes involved. This lack of phase information can explain why the imaginary part of the $\rho_{SE}$ is zero even though that of the $\rho_{QST}$ is not. This can be addressed by measuring the JSP experimentally as in~\cite{thekkadath2022measuring,garay2023fiber,hurvitz2024phase}, which will determine the phase of a corresponding FWM process through $c_{j}\propto f_{j}\propto e^{i(JSP)}$ (see Eq.~\ref{eq:psi_si-coeff} and Supplement~1 for more details). Third, we assume the distinguishabilities that can undermine the transverse-mode entanglement do not exist outside the spatial and spectral DOFs. Although the linear polarizers and a cross-spliced fiber are employed to compensate any residual polarization and temporal distinguishabilities in the system (the second PMF in the cross-spliced PMF corrects for polarization- and transverse-mode parity-dependent temporal walk-offs introduced in the first PMF~\cite{meyer2013generating}), there is a chance that some unaccounted distinguishabilities still remain. As a remedy, extending the technique to other DOFs, e.g., polarization and time-resolved characterization, may be helpful~\cite{fang2016multidimensional,ploschner2022spatial}. Alternatively, we may also consider a full frequency-resolved transverse-mode stimulated emission \emph{tomography}, whose projection measurements naturally accounts for all possible distinguishabilities. This may produce better estimation, albeit will lack the information on the source of distinguishability if it exists outside the spatial and spectral domains. Ultimately, all the resolutions presented here can lead to potential improvements in the numerical model used in previous works~\cite{cruz2016fiber,garay2016photon,cruz2014configurable} to better predict the quantum state.

\section{Conclusion} \label{sec:Conclusion}
We have applied a stimulated-emission-based characterization technique to reveal the transverse-mode-frequency relation of photon pairs created from four-wave mixing processes in few-mode PMF. We measured the joint spectral intensities and transverse modes of the stimulated signal while controlling the pump and seed transverse modes and the seed wavelength. From these measurement results, we identified FWM processes predicted by theory and an additional parity birefringence dispersion parameter $\delta$ required to explain the spectral distinguishability between $\ket{e_{s}e_{i}}$ and $\ket{o_{s}o_{i}}$ photon-pair states. We demonstrated the efficiency of our technique by comparing with spontaneous measurements for imaging signal transverse modes.
Leveraging the efficiency of stimulated-emission-based measurement, we demonstrated real-time imaging capability and investigated the quantum properties of a transverse-mode entangled photon-pair source. We illustrated how the transverse-mode quantum state of the photon pairs can be estimated from the spatio-spectral measurements, specifically transverse-mode-resolved stimulated JSIs. The estimated density matrix $\rho_{SE}$ showed qualitative agreement with that measured from a standard transverse-mode QST, $\rho_{QST}$. Estimating $\rho_{SE}$ provided additional information on spectral distinguishability that can lead to low transverse-mode entanglement.

This stimulated-emission-based spatio-spectral characterization technique may benefit from a possible extension to a spectrally-resolved transverse-mode stimulated-emission tomography~\cite{liscidini2013stimulated,rozema2015characterizing,fang2016multidimensional} and joint spectral phase measurement~\cite{thekkadath2022measuring,garay2023fiber}. By allowing the diagnosis of potential causes of entanglement degradation originating from other degrees of freedom, this method may be utilized to create versatile fiber-based photon-pair sources with entanglement in frequency and transverse mode~\cite{kim2023generating,kim2023towards}, as well as transverse-mode-frequency hybrid-entanglement~\cite{cruz2016fiber}. We anticipate this stimulated-emission characterization technique may also be extended beyond few-mode optical fiber to efficiently diagnose and optimize photon-pair sources in a variety of quantum systems with high dimensionality~\cite{bozinovic2013terabit,ramachandran2015scalability,defienne2016two,shahar2023photon,onodera2024scaling} and different degrees of freedom~\cite{fang2016multidimensional,barreiro2005generation,serino2024orchestrating,presutti2024highly}.

\begin{backmatter}
\bmsection{Funding}
National Science Foundation (1806572, 1640968, 1839177, 2207822); U.S. Department of Energy, Office of Science, Biological and Environmental Research program (Award No. DE-SC0023167).
\bmsection{Acknowledgments}
We thank Joel Carpenter for help with transverse-mode control using an SLM, Elizabeth Goldschmidt for help with aligning the CW dye laser, Oliver Wang and Xinan Chen for the initial experimental setup, Offir Cohen and Bin Fang for helpful discussion about density matrix error analysis, Xiao Liu, Daniel Shahar, and Siddharth Ramachandran for fruitful discussion about OAM modes, and Soho Shim for valuable discussion during manuscript preparation.
\bmsection{Disclosures}
The authors declare no conflicts of interest.
\bmsection{Data Availability Statement}
Data underlying the results presented in this paper are
not publicly available at this time but may be obtained from the authors upon reasonable request.
\bmsection{Supplemental document}
See Supplement~1 for supporting content. 
\end{backmatter}

\bibliography{bib-fiber,bib-dbkpublications,bib-quombis}
\end{document}


\maketitle

\section{Detailed quantum state representations}
\emph{Quantum state of the pump},
\begin{equation} \label{eq:psi_p}
    \ket{\psi_{p}}=\sum_{j}\int d\omega_{p}\,a_{j}\ket{\omega_{p},T_{p},x,...}_{j}=\sum_{j}A_{j}\otimes\ket{T_{p}}_{j},
\end{equation}
\begin{equation} \label{eq:psi_p1p2}
    \ket{\psi_{p_{1}p_{2}}}=\ket{\psi_{p}}^{\otimes2}=\sum_{j}\int d\omega_{p}^2\,b_{j}\ket{\omega_{p}\omega_{p},T_{p_1}T_{p_2},xx,...}_{j}=\sum_{j}B_{j}\otimes\ket{T_{p_1}T_{p_2}}_{j}
\end{equation}
where the prefactors satisfying the pump condition $\ket{\psi_p}=\ket{\psi_{p_1}}=\ket{\psi_{p_2}}$ (given that we do not have individual control over the two pumps) are
\begin{equation} \label{eq:psi_p-coeff}
    a_{j}=M_{aj}\alpha(\omega_{p}),\quad A_{j}=\int d\omega_{p}\,a_{j}\ket{\omega_{p},x,...}_{j},
\end{equation}
\begin{equation} \label{eq:psi_p1p2-coeff}
    b_{j}=M_{bj}\alpha^2(\omega_{p})=a_{j}^2,\quad B_{j}=\int d\omega_{p}^2\,b_{j}\ket{\omega_{p}\omega_{p},xx,...}_{j}
\end{equation}
Here, $M_{aj}$ and $M_{bj}$ are (complex) coefficients that depend on the relative amplitude and phase of different pump (transverse) modes, satisfying $\braket{\psi_{p_{1}p_{2}}|\psi_{p_{1}p_{2}}}=1$ and $\braket{\psi_{p}|\psi_{p}}=1$.

\emph{Quantum state of the signal-idler photon pair},
\begin{equation} \label{eq:psi_si-suppl}
    \ket{\psi_{si}}=\sum_{j}^{N}\int d\omega_{s}\,d\omega_{i}\,c_{j}\ket{\omega_{s}\omega_{i},T_{s}T_{i},yy,...}_{j}=\sum_{j}^{N}C_{j}\otimes\ket{T_{s}T_{i}}_{j},
\end{equation}
where the prefactors are
\begin{equation} \label{eq:psi_si-coeff-suppl}
    c_{j}=M_{cj}b_{j}f_{j}(\omega_{s},\omega_{i})O_{j}(T_{p_1},T_{p_2},T_{s},T_{i}),\ C_{j}=\int d\omega_{s}\,d\omega_{i}\,c_{j}\ket{\omega_{s}\omega_{i},yy,...}_{j}.
\end{equation}
Notice that $\sqrt{P_{p_{1}j}P_{p_{2}j}}$ in Eq.~3 in the main text, which varies with the transverse modes and frequencies involved in the process $j$, is replaced with $b_{j}$ to more accurately reflect the pump contribution.

\emph{Joint spectral amplitude (JSA)},
\begin{equation} \label{eq:f-JSA-suppl}
    f_{j}(\omega_{s},\omega_{i})=\left|f_{j}(\omega_{s},\omega_{i})\right|e^{i\arg\{f_{j}(\omega_{s},\omega_{i})\}}=\sqrt{\mathrm{JSI}}e^{i\left(\mathrm{JSP}\right)},
\end{equation}
where the definitions, joint spectral intensity $\mathrm{JSI}=|f_{j}(\omega_{s},\omega_{i})|^2$ and joint spectral phase $\mathrm{JSP}=\arg\{f_{j}(\omega_{s},\omega_{i})\}$, are used.

\emph{Transverse-mode overlap integral},
\begin{equation} \label{eq:o-overlap-suppl}
    O_{j}(T_{p_1},T_{p_2},T_{s},T_{i})=M_{oj}\int d^{2}\vec{r}\,T_{p_1}(\vec{r})T_{p_2}(\vec{r})T_{s}^{*}(\vec{r})T_{i}^{*}(\vec{r}),
\end{equation}
where $M_{oj}$ is the normalization constant satisfying $\sum_{j}|O_{j}|^2=1$, $\vec{r}$ the transverse position, and $T_{\nu}(\vec{r})$ the transverse electric field of photon $\nu$.

\section{Seed laser calibration}
\begin{figure}[ht]
    \centering
    \includegraphics[width=0.7\textwidth]{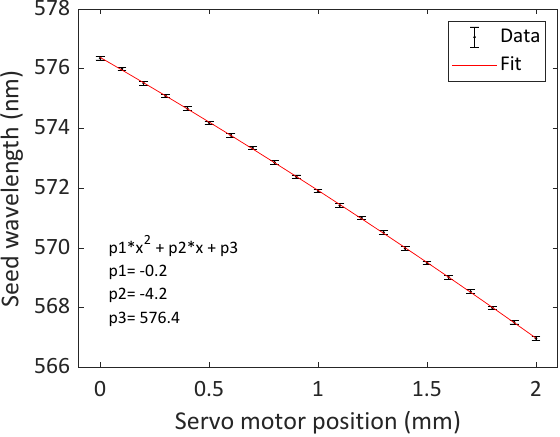}
    \caption{Calibration curve of the dye (seed) laser in servo motor steps of 0.1 mm specified by a quadratic fit. The seed is calibrated by measuring its center wavelength as a function of the DC servo motor (Thorlabs Z812B) position, which is responsible for the wavelength tuning (rotates the birefringent filters in the laser cavity) of the dye laser.}
    \label{fig:SeedCalibration}
\end{figure}
\newpage

\section{Transverse-mode overlap integral and JSI intensity}
We experimentally confirmed that the transverse-mode overlap integral factor $O_{j}(T_{p_1},T_{p_2},T_{s},T_{i})$ in Eq.~\ref{eq:psi_si-coeff-suppl} can explain the relative intensities of the FWM processes observed in Fig.~3 in the main text as predicted by the theoretical model~\cite{garay2016photon}. As shown in Eq.~\ref{eq:psi_si-coeff-suppl}, this overlap integral $O_{j}$ affects the photon-pair generation efficiency of a FWM process $j$. $|O_{j}|^2$, proportional to the relative intensity of process $j$, is maximal ($|O|^2\approx0.35$) when all the participating transverse modes are identical, and is smaller by a factor of about 2.2 ($|O|^2\approx0.15$) when there are two $\ket{e}$'s and two $\ket{o}$'s (see Eq.~\ref{eq:o-overlap-suppl}). This indeed explains why the FWM processes in Fig.~3 at B $(o,o,o,o)$ and C $(e,e,e,e)$ appear brighter compared to those at A $(e,o,o,e)$, and D $(e,o,e,o)$. Additional asymmetry in intensity between the processes at B and C (see Fig.~3(c)) can be explained by the unaccounted pump ($\sqrt{P_{p_1}P_{p_2}}$ in Eq.~\ref{eq:psi_si-coeff-suppl}) or the seed power difference for the $\ket{e}$ and $\ket{o}$ modes across the wavelength range.

\section{Transverse mode imaging method comparison}
To investigate the efficiency of stimulated-emission-based imaging, we compare the transverse-mode images of the signal with and without the seed, i.e., using stimulated and spontaneous FWM. This is achieved by simply blocking or unblocking the beam path of the seed in our experimental setup (see Fig.~1 in the main text).

\begin{figure}[ht!]
    \centering
    \includegraphics[width=\textwidth]{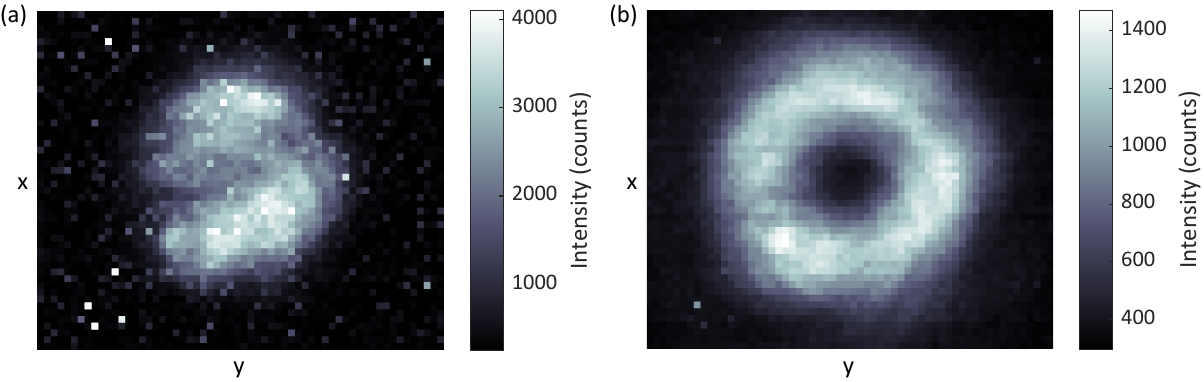}
    \caption{Transverse-mode images of the signal from (a) spontaneous (without seed) FWM with an exposure time of 14235 ms and (b) stimulated FWM (with $\ket{d}$ seed) with an exposure time of 400 ms. For both cases, $\ket{d}$ pump and broad signal spectral filters are used, transmitting all 4 A-D FWM processes. The stimulated signal image (b) corresponds to a vertical slice of the Fig.~3(c) JSI plot at BC. The intensities of the images are measured by the camera.}
    \label{fig:SpontVsStim}
\end{figure}

Figure~\ref{fig:SpontVsStim}(a) shows a transverse-mode image of signal photons from spontaneous FWM taken with an exposure time of 14,235 ms, the longest available with our CMOS camera. With a shorter exposure time of 400 ms, it does not exhibit any noticeable spatial structure above the background counts. With this same shorter exposure time of 400 ms ($\approx$ 36 times smaller than that used in the spontaneous case), the stimulated image in Fig.~\ref{fig:SpontVsStim}(b) shows a much clearer mode structure with less background noise (see Sec.~\ref{sec:donut} for discussion on the \emph{donut} shape and its relationship with coherence). We attribute the noise in the spontaneous case to Raman scattering, which can be a dominant feature in the SFWM processes with higher-order transverse modes~\cite{smith2009photon}. With the experimental parameters used here, the Raman noise and FWM signal spectrally overlap and have comparable spectral strength. To identify the underlying cause for future quantum applications, we can employ~\cite{garay2023fiber} idler transverse-mode imaging with potentially less Raman contamination than that of the signal~\cite{smith2009photon}, pixel-by-pixel coincidence counting~\cite{cruz2016fiber,zia2023interferometric}, or ghost imaging-type measurements~\cite{gregory2020imaging,aspden2015photon,moreau2018ghost}. Meanwhile, stimulated-emission-based imaging allows us to investigate the spatial mode of the signal photons independent of such noise processes with higher contrast. This is consistent with previous studies~\cite{liscidini2013stimulated,fang2014fast} that showed stimulated-emission-based characterization enabling more efficient measurement of single-photon-level states. With this efficiency, remarkably, real-time monitoring of the signal transverse mode is possible (see Visualization~1 data sequence as the seed wavelength is varied), similar to~\cite{oliveira2019real} with spontaneous parametric down-conversion crystals in free-space.

Moreover, stimulated-emission imaging is advantageous in resolving and identifying the seemingly intermingled multi-dimensional FWM processes, especially as exhibited in our few-mode PMF. Directly imaging a spontaneously emitted signal yields a single image as shown in Fig.~\ref{fig:SpontVsStim}(a). If more than one exists within the spectral window of interest, it is hard to distinguish the individual contributions from different FWM processes. While it is possible to spectrally resolve these processes and transverse modes with narrow-band spectral filters as in~\cite{cruz2016fiber,zia2023interferometric}, this entails multiple complications: longer measurement times, additional equipment for single-photon-level transverse-mode measurement, and coincidence counting to overcome the low contrast. With stimulated imaging, different signal transverse modes can be imaged through a controlled choice of the seed transverse mode and wavelength (see Fig.~2 in the main text), providing insight into the transverse-mode-spectral relationship of the photon pairs created in the PMF.

\section{Observation of donut mode and coherence} \label{sec:donut}
When the signal spectral filtering window is sufficiently widened to let all of the 4 FWM processes through, the stimulated transverse mode image obtained at the seed wavelength of $\lambda_d=570.8\mathrm{\ nm}$ as shown in Fig.~\ref{fig:SpontVsStim}(b) -- denoted as BC (in between the processes B and C) in our Visualization~2 data sequence, Fig.~\ref{vid:SET-Tmodes,JSI,animated}, and  -- shows apparently mixed behavior: a \textit{donut} mode (intensity null at the center). This is often associated with an orbital angular momentum mode~\cite{mair2001entanglement,krenn2017orbital}, or a superposition mode $\ket{r,l}=(\ket{e}\pm i\ket{o})/\sqrt{2}$ when expressed in the LP mode basis (see Fig.~1). Coincidentally, an incoherent mixture (a mixed state) of the $\ket{e}$ and $\ket{o}$ modes, $(\ketbra{e}{e}+\ketbra{o}{o})/2$ exhibits an identical intensity distribution as the superposition states $\ket{r}$ and $\ket{l}$. Thus, without sufficient spectral filtering, solely imaging the stimulated transverse mode at BC with a $\ket{d}$ seed may not give enough information to solve the ambiguity. The additional information provided by the transverse-mode-resolved JSI measurements of Fig.~3(a-b) can be used to deduce that the donut mode is from an incoherent mixture of $\ket{o}$ and $\ket{e}$. This is a result of simultaneously exciting two separate spectrally distinguishable FWM processes, B and C, at $\lambda_{d}=570.8\mathrm{\ nm}$. With stimulated emission, we are also able to monitor the stimulated transverse mode while changing the phase of the pump or seed beam in real time (see Visualization~1 data sequence) to assess the coherence of the involved FWM processes.

\section{Real-time and animated videos}
\begin{figure}[ht!]
    \centering  
    \includegraphics[width=0.5\textwidth]{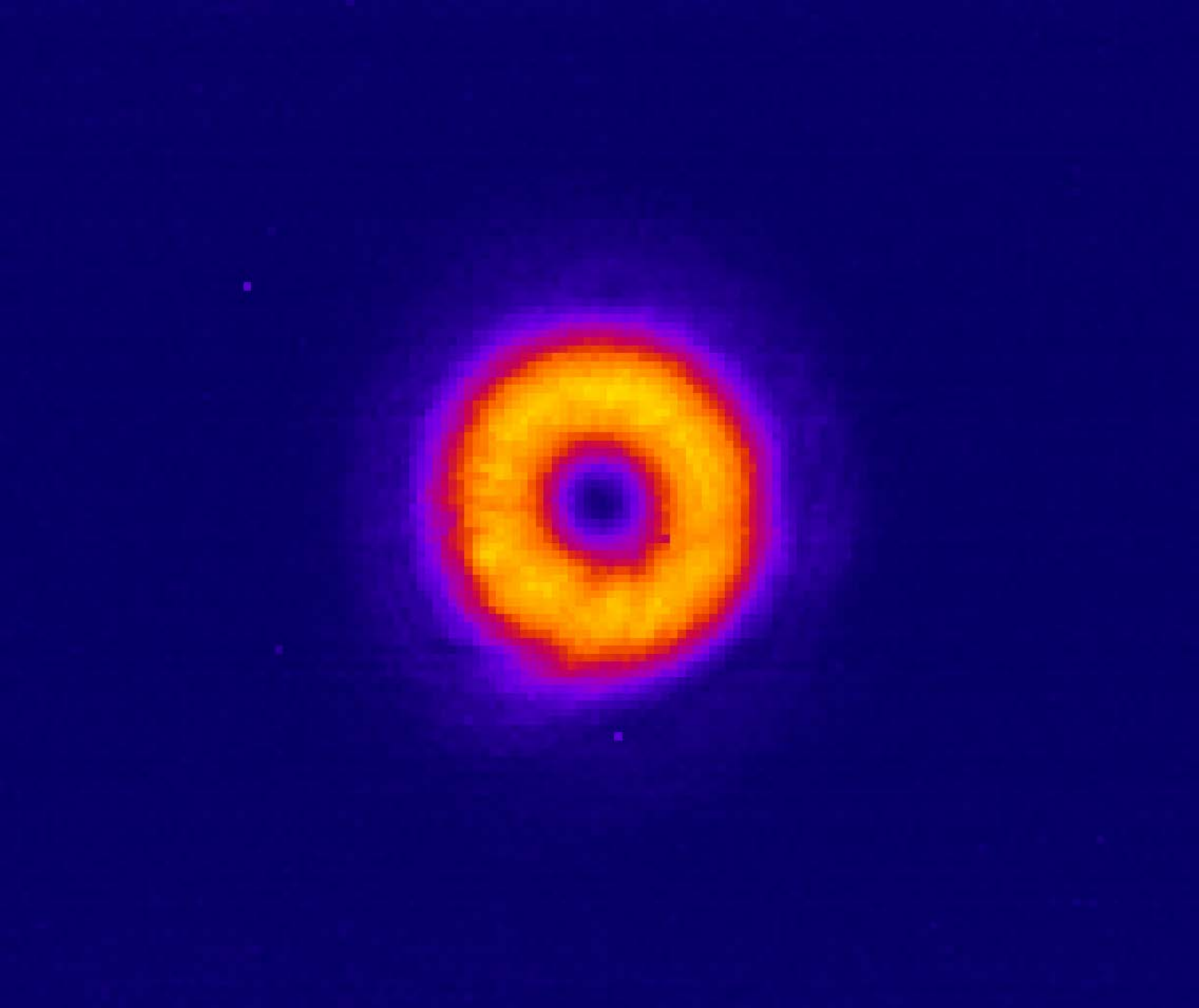}
    \caption{Screenshot of a real-time video Visualization~1. This raw video data shows in real-time how the transverse mode of the stimulated beam changes as the seed wavelength (servo motor) is scanned. Exposure time for this video is kept the same as for Fig.~\ref{vid:SET-Tmodes,JSI,animated}, 400 ms. A difference is that this video shows a 152 px $\times$ 128 px full field of view, while others (Figs.~3 and 5 in the main text, and Fig.~\ref{vid:SET-Tmodes,JSI,animated}) are trimmed to the 60 px $\times$ 50 px region of interest.}
    \label{vid:SET-Tmodes,JSI,realtime}
\end{figure}

\begin{figure}[ht!]
    \centering
    \includegraphics[width=\textwidth]{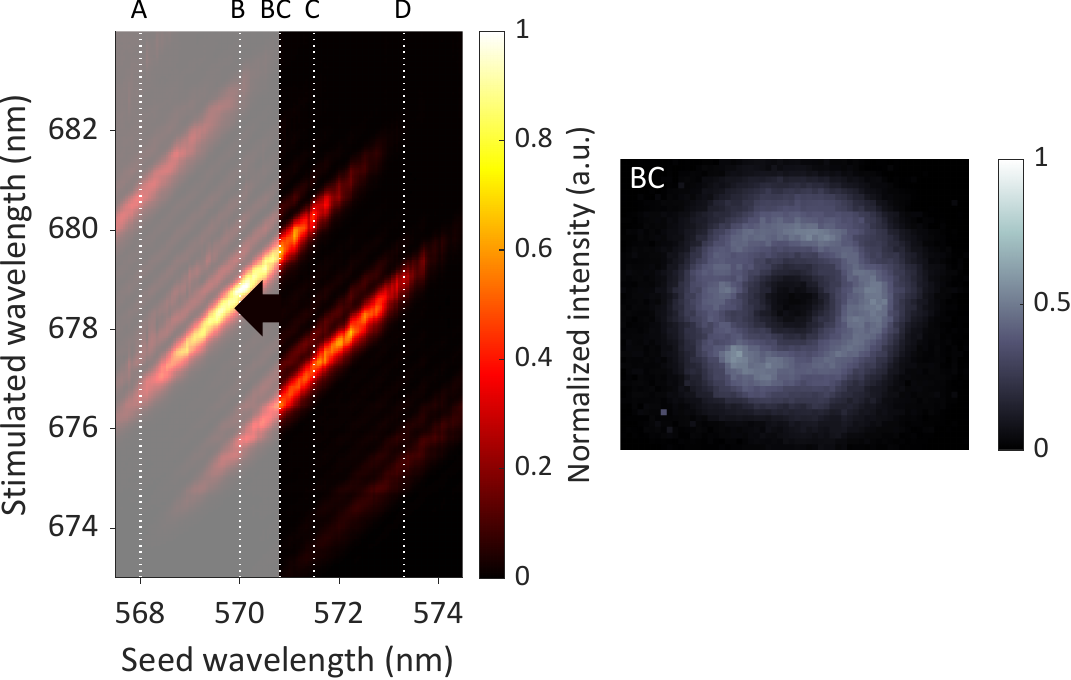}
    \caption{Screenshot of an animated video Visualization~2 illustrating the continuous evolution of the stimulated signal transverse mode as the seed wavelength is scanned. The JSI is horizontally scanned with the seed wavelength (left) as the corresponding stimulated image is shown simultaneously (right). The center seed (idler) wavelengths and the corresponding stimulated (signal) transverse mode images are labeled with the FWM processes (A-D). Notice that BC is located in the middle of the two processes, B and C. This animation is recreated from two separate measurements of a JSI and individual stimulated beam images each taken with 400 ms exposure time. The images are background subtracted and normalized in intensity.}
    \label{vid:SET-Tmodes,JSI,animated}
\end{figure}

\newpage
\section{Transverse-mode quantum state tomography setup}
\begin{figure}[ht!]
    \centering  
    \includegraphics[width=\linewidth]{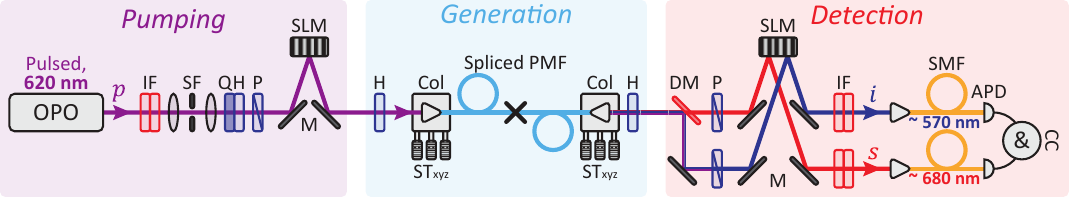}
    \caption{Experimental setup for transverse-mode quantum state tomography. Signal-idler photon pairs are created from $2.5\mathrm{\ cm}\times2$ cross-spliced PMF and projected on to specific transverse-mode basis states by an SLM and a single-mode fiber. The projected photons are coincidence counted by APDs and a coincidence counter. $p$: pump, $s$: signal, $i$: idler, IF: interference filter, SF: spatial filter consisting of a pinhole and a convex lens pair, Q: quarter-wave plate, H: half-wave plate, P: linear polarizer, M: mirror, SLM: spatial light modulator, DM: dichroic mirror, Col: collimator, $\mathrm{ST_{xyz}}$: xyz-translation stage, APD: avalanche-photodiode, CC: coincidence counter.}
    \label{fig:QSTsetup}
\end{figure}

Experimental setup for quantum state tomography (QST)~\cite{altepeter2005photonic} in the transverse-mode basis is shown in Fig.~\ref{fig:QSTsetup}. At the \emph{detection} part in the Fig.~2 experimental setup, we install an additional SLM and two pairs of a single-mode fiber and a single-photon detector (APD, Excelitas SPCM-AQ4C) that act together as a projection measurement device for each signal and idler photon. The two APDs are connected to a time tagger (IDQ ID800 time-to-digital converter) for coincidence counting. Using this setup, we project the signal-idler transverse-mode states into six mutually unbiased measurement basis states ($e, o, d, a, r, l$) and conduct 36 coincidence measurements ($ee, eo, ..., lr, ll$)~\cite{altepeter2005photonic,mair2001entanglement,agnew2011tomography,langford2004measuring,hiekkamaki2019near,kim2023towards}. From the 36 coincidence count results, we find a density matrix that best explains (maximum likelihood) the quantum state measured~\cite{altepeter2005photonic}.

\bibliography{bib-fiber,bib-dbkpublications,bib-quombis}